\documentclass[1p]{elsarticle}
\usepackage{amsmath,amssymb}
\usepackage{amsthm}
\usepackage{epsfig}
\usepackage{graphicx}
\usepackage{float}
\begin{document}
\begin{frontmatter}
\title{Non-Markovian evolution of a two-level system 
interacting with a fluctuating classical field via dipole interaction}
\author[1]{Samaneh Hesabi} 
\author[1,2]{Davood Afshar\corref{cor1}}
\ead{da\_afshar@yahoo.com}
\author [3]{Matteo G. A. Paris\corref{cor1}}
\cortext[cor1]{Corresponding authors}
\ead{matteo.paris@fisica.unimi.it}
\address[1]{Department of Physics, Faculty of Science, Shahid Chamran University of Ahvaz, Ahvaz, Iran}
\address[2]{Center for research on Laser and Plasma, Shahid Chamran University of Ahvaz, Ahvaz, Iran }
\address [3]{Quantum Technology Lab, Dipartimento di Fisica {\em Aldo Pontremoli} \\  Universit\`a degli Studi di Milano, I-20133 Milano, Italy}
\begin{abstract}
We address memory effects in the dynamics of a two-level 
open quantum system interacting with a classical fluctuating 
field via dipole interaction. In particular, we study the 
backflow of information for a 
field with a Lorentzian spectrum, and reveal the existence of 
two working regimes, where memory effects are governed either
by the energy gap of the two-level system, or by the 
interaction energy. Our results shows that non-Markovianity 
increases with time, at variance with the results obtained for dephasing 
and despite the dissipative nature of the 
interaction, thus suggesting that the corresponding memory effects
might be observed in practical scenarios.
\end{abstract}
\begin{keyword}
non-Markovianity \sep memory effects \sep stochastic electromagnetic field
\end{keyword}
\end{frontmatter}
\section{Introduction}
\label{Intro}
The dynamics of a closed quantum system is reversible and the 
time evolution of its states, which is governed by the Schr\"odinger 
equation, is described by unitary maps. On the other hand, 
for an open quantum system interacting with its environment, the 
state evolution is no longer reversible and is described by completely-positive, trace-preserving (CPTP) maps, which themselves result from 
the partial trace of the (unitary) joint evolution of the open system 
$+$ environment. In turn, the evolution equation for an open quantum 
system, the so-called Master equation (ME), should be derived from 
the Schr\"odinger equation of the overall system, upon tracing out the degree of freedom of the environment \cite{ref1,ref2}. 
\par
In most cases, obtaining a ME is quite challenging and, in turn, 
there are only few examples of open quantum systems for which 
an exact ME may be derived. In the other cases, the most common approximations used to derive the ME of an open quantum systems 
are those referred to as the {\it Born} and {\it Markov} 
approximations \cite{ref1}. 
The Born approximation amounts to assume a weak coupling between the 
open system and its environment. Markov approximation consists instead 
in assuming that the dynamics of the environment is slow compared to 
the system's one. In the Markovian regime, the state of the system 
at time $t$ is independent
on its past, i.e. we do not need to go
backwards in time to account for memory effects. In turn, the loss 
of memory effects corresponds to CPTP maps of the Lindblad type 
\cite{ref4}, satisfying the so-called divisibility property. 
\par 
Lindblad-Markov MEs are valid tools in describing the dynamics of several 
systems, but they are unable to describe coherent phenomena occurring 
in several solid-state and biological systems \cite{ref7,ref8,ref9}, 
as well as in material systems with a photonic band gap \cite{ref5,ref6}.
If memory effects cannot be neglected, the dynamics is referred to as 
non-Markovian. Non-Markovianity may also cause a backflow of information from the environment to the system, and this may be exploited e.g. 
to enhance security of quantum key distribution \cite{vas1} or
outperform metrological strategies based on uncorrelated states 
\cite{add2}. We also remind that there are MEs suitable to describe 
the dynamics of non-Markovian open quantum systems as for example 
the Nakajima-Zwanzig ME \cite{ref10,ref11} or the time-convolutionless 
one \cite{ref12,ref13}. 
\par
In order to characterize and quantify non-Markovianity, several figures of merit have been introduced. A measure has been proposed 
(BLP) on the basis of memory effect \cite{ref14} and another one 
(RHP) in terms of entanglement of the system and its environment 
\cite{ref15}. In addition, other measures have been proposed, based on 
quantum Fisher information \cite{ref16}, mutual information \cite{ref17} and temporal steering \cite{ref18}. In this paper, since we are interested in 
discussing the sources of memory effects for an atom interacting with
its electromagnetic environment, we stick with the original BLP
definition in terms of backflow of information.
\par
A two-level atom interacting with its electromagnetic environment corresponds
to an open quantum system \cite{ref19,ref20,ref21,ref22,ref23,ref24,ref25}. A well-known and solvable model to describe 
its dynamics is the Jaynes-Cummings one, describing the interaction with 
a single-mode in terms of the dipole moment \cite{ref26,ref27,ref28}. 
In recent years, some studies have been performed 
about the non-Markovianity of this system in the presence of various environments. In particular, 
non-Markovianity has been considered for a two-level system coupled 
to a single mode of the field via the Jaynes-Cummings model using BLP measure \cite{ref26} and coupled to random external fields using 
different measures \cite{ref33}. The non-Markovianity of a 
damped Jaynes-Cummings model \cite{ref30,ref34}
and for the coupling to a quantized bosonic field \cite{ref19} has 
been studied as well. 
\par
If a two-level system is exposed to a fluctuating field
with a broad spectrum, the equations of motion are not linear and the 
response of the system may not obtained easily. In those situations, 
it is convenient to employ semiclassical stochastic methods 
\cite{ref35,ref36,oli12,nmce,qpFGN} to address the time 
evolution. Since this
is a common situation of practical interest for various 
applications, we here investigate the non-Markovianity of a two-level system interacting with a stochastic field with a Lorentzian spectrum. 
\par
The paper is structured as follows. In Section 2 we 
describe our model, solve the equations of motion for the specific case
of a Lorenzian spectrum and introduce the BLP measure of non-Markovianity. 
Our results and the different working regimes are then illustrated in 
Section 3. Section 4 closes the paper with some concluding remarks.
\section{The system and the interaction model} 
We consider a two-level system interacting with a classical, possibly fluctuating, e.m. field. The dynamics of the system is described by the 
Hamiltonian 
\begin{equation}
H = {H_0} - {\boldsymbol \mu} \cdot {\boldsymbol E},
\label{E2}
\end{equation}                                         
where $H_0=\frac12\, \Omega\, \sigma_3$ is the free Hamiltonian of the 
two-level system, ${\boldsymbol \mu}= \mu \cos\alpha\, \sigma_1$ is the dipole 
moment of the system, and $ {\boldsymbol E} 
\equiv {\boldsymbol E}_t$ is an external, possibly fluctuating, electromagnetic field, 
$\alpha$ is the angle between the dipole moment and the field.  
The field may be a deterministic function of time, describing a 
driving field, or a stochastic processing, describing a fluctuating background field. The latter situation is that of interest in this 
work.
The state of the system at the time $t$ is given by 
\begin{align}
\rho_t = \mathbb{E}\Big[|\psi_t \rangle\langle\psi_t|\Big]_E\,,
\label{rhot}
\end{align}
where $\mathbb{E}[\cdots]_E$ denotes the average over the different
realizations of the background field, intended as a stochastic process.
The single-realization state of the system is given by
$| \psi_t 
\rangle  = a_t | 
0 \rangle  + b_t|1\rangle$, where $|j\rangle$, $j=0,1$ are the 
eigenstates of the free Hamiltonian, ${H_0}|j\rangle  = {E_j} |j\rangle$, with ${\left| a_t \right|^2} + {\left| b_t \right|^2} = 1$, $\forall t$.
The instantaneous dipole moment and energy of the system are 
given by 
\begin{align}
M_t & = \mu (a_t b_t^* + a_t^* b_t), \\
W_t & = \frac12  \Omega\, ({\left| b_t \right|^2} - {\left| a_t 
\right|^2})\,,
\end{align}                                                           
where we already employed natural units, i.e. $\hbar=1$.                                                                                                                                                       
In turn, the Schr\"odinger equation for the two-level systems 
may be written in terms 
of $M_t$ and $W_t$ as follows
\begin{align}
\ddot M_t + \Omega^2M_t & =  - \left(2\mu\cos\alpha\right)^2\, 
W_t\, E_t, \\
\dot W_t + \beta_s (W_t + \frac12 \Omega) &= \dot M_t E_t \,,
\label{E8}
\end{align}
where $\beta_s$ is the 
Einstein coefficient for spontaneous emission. 
The solution of the above equations is given by \cite{ref35,ref36}: 
\begin{align}
M_t =& M_0\cos \Omega t + \dot M_0\, \frac{\sin \Omega t}{\Omega} 
- \frac{\left(2\mu\cos\alpha\right)^2}{\omega}
\int_0^t\!\!ds\,\sin[ \Omega (t - s)]\, W_s\,E_s, \\
W_t =& - \frac12 \Omega  + (W_0 + \frac12 \Omega)\, e^{-\beta_st} 
 + \int_0^t\!\! ds\, W_s\,G_{ts} \notag \\
 & - M_0 \int_0^t\!\! ds\,\sin \Omega s\, e^{-\beta_s(t-s)} E_s
 + \dot M_0\int_0^t\!\!ds\, \cos \Omega s\, \, e^{-\beta_s(t-s)} E_s,
\label{E9}
\end{align} 
where 
\begin{equation}
G_{ts} =  - \left(2\mu\cos\alpha\right)^2 e^{-\beta_s(t-s)} 
\int_s^t\!\! dy\,E_y\,E_s \cos \Omega (s-y)\,,
\label{E10}
\end{equation}  
and $M_0$, $W_0$ denotes the initial values of the dipole moment and the energy, respectively. In the following, we assume that the field amplitude fluctuates around a vanishing average $\mathbb{E}[E_t]_E = 0$. 
In this case, we have
\begin{align}
\mathbb{E}[M_t]_E
& = M_0 \cos \Omega t, \\
\mathbb{E}[W_t]_E & = \frac{1}{2\pi i}\oint dz\, \frac{e^{zt}
\Big[z\,W_0 - \frac12 \beta_s \Omega\Big]}{z\left[z + \beta_s + \left(2\mu\cos\alpha\right)^2\,F(z)\right]} \label{E11} \\ 
F(z) & = \frac{z}{2}\int_{- \infty}^\infty  
\!\!\!\! d\omega\, {\frac{{I(\omega ) }}{{{{(\omega  - \Omega )}^2} + {z^2}}}}\,,\notag
\end{align}                                                                                                                                                                             where $I(\omega )$ is the power spectrum of field, i.e. the Fourier transform of its correlations function $\mathbb{E}[E(t)E(s)]_E$.
On the assumption that the background radiation field has a 
Lorentzian spectrum centered at the resonance $\Omega$, i.e.
\begin{equation}
I(\omega ) = \frac{{{I_0}{\beta ^2}}}{{[{{(\omega  - \Omega )}^2} + {\beta ^2}]}},
\label{E13}
\end{equation}
$\mathbb{E}[W_t]_E$ may be simplified to
\begin{align}
\mathbb{E}[W_t]_E = A  \Big[-1 + e^{-\gamma t} \cos \lambda t\Big]+ W_0\, e^{- \gamma t}\cos \lambda t
+ \frac12\,e^{- \gamma t}\, \beta_s\,\frac{\sin \lambda t}{\lambda} (B-A),
\label{E14}
\end{align}
where
\begin{align}
A & = \frac{\Omega \beta_s}{2 \beta_s +  
\pi I_0 \left(2\mu\cos\alpha\right)^2}, 
& B    = \frac{\Omega \big[\beta-\pi I_0  
\left(2\mu\cos\alpha\right)^2\big]}{2 \beta_s +  
\pi, I_0 \left(2\mu\cos\alpha\right)^2},\\ 
\gamma  & = \frac12 (\beta+\beta_s), 
&  \lambda^2    = \frac12 \left(2\mu\cos\alpha\right)^2 
\pi \beta I_0 - \frac14 (\beta-\beta_s)^2.
 \label{E15}
\end{align}
Using Bloch representation in terms of the Pauli matrices, 
the evolved state of the system, i.e. $\rho_t$ in Eq. (\ref{rhot}),
may thus rewritten as 
\begin{align}
\rho_t = \frac12 \Big(\mathbb{I} + \mathbb{E}[W_t]_E\, \sigma_3 
+ \mathbb{E}[M_t]_E \,\sigma_1
\Big) \label{arhot}
\,,
\end{align}
which says that the dynamics of coherence does depend only on 
its energy gap $\Omega$, whereas the populations are affected also 
by the variables governing the interaction Hamiltonians. The purity
of the state at time $t$ is given by 
\begin{align}
\mu_t = \hbox{Tr}[\rho_t^2] = \frac12 \left(1+\mathbb{E}[M_t]_E^2 +\mathbb{E}[W_t]_E^2  \right)\,.
\end{align}
\subsection{Quantification of memory effects}
Before addressing the dynamics of non-Markovianity in details, let us 
briefly review how memory effects (due to non-Markovian 
dynamics) may be quantified using the time dependence of the trace 
distance $D(\rho_{1t},\rho_{2t}) = \frac{1}{2}\left\| 
\rho_{1t} - \rho_{2t} \right\|$ between a pair 
of evolved states of the system \cite{ref14}. In the previous formula,
$\left\| A \right\|$ denotes the trace-norm of the operator $A$ 
i.e.  $\left\| A \right\| = \hbox{Tr}[\left| A \right|] = \hbox{Tr}[\sqrt {{A^\dag }A}] = \sum_k \sqrt{|a_k|}$, $a_k$ being the eigenvalues 
of $A^\dag A$. 
As a starting point, we remind that any
completely-positive and trace-preserving map ${\cal E}_t$ 
is also {\em contractive}, i.e. the trace distance between 
any two evolved states decreases
$D(\rho_{1t},\rho_{2t}) < D(\rho_{10},\rho_{20})$, 
where $\rho_t={\cal E}_t [\rho_0]$. For any Markovian process,
the divisibility property makes the contractivity properties to hold 
for any two chosen values of time, i.e. 
\begin{equation}
D(\rho_{1t},\rho_{2t}) < D(\rho_{1s},\rho_{2s}) \qquad
\forall t > s
\label{E19}\,.
\end{equation}
In turn, this monotonic decrease of distinguishability may be 
understood as an irreversible flow of information 
from the system to the environment. On the other hand, for a 
non-Markovian process, divisibility is lost, and 
the trace distance may increase in some time-interval. 
This means that in those cases information is {\em flowing back} 
from the environment to the system. This property may be employed
to define a measure of non-Markovianity as follows:  
\begin{equation}
N_T = {\max }_{\rho_{10},\rho_{20}} \int_{\sigma  > 0}^T\!\!\!\! 
ds\,\sigma (s,\rho_{1s},\rho_{2s}),
\label{E20}
\end{equation}
where $\sigma (t,\rho_{1t},\rho_{2t})$ is the rate of change
of the trace distance
\begin{equation}
\sigma (t,\rho_{1t},\rho_{2t}) = 
\frac{d}{{dt}}\,D(\rho_{1t},\rho_{2t})\,,
\label{E21}
\end{equation}
and the maximisation is performed over all the possible pairs of {\em initial} states. The integration is meant over the intervals where 
$\sigma$ is positive up to a maximum time $T$, which corresponds to a
maximum interaction time, e.g. due to a finite observation time or, in case
of a propagating system, due to the finite size of the region with
nonzero field.
\section{Dynamics of non-Markovianity}
\begin{figure}[h!]
\includegraphics[width=0.32\columnwidth]{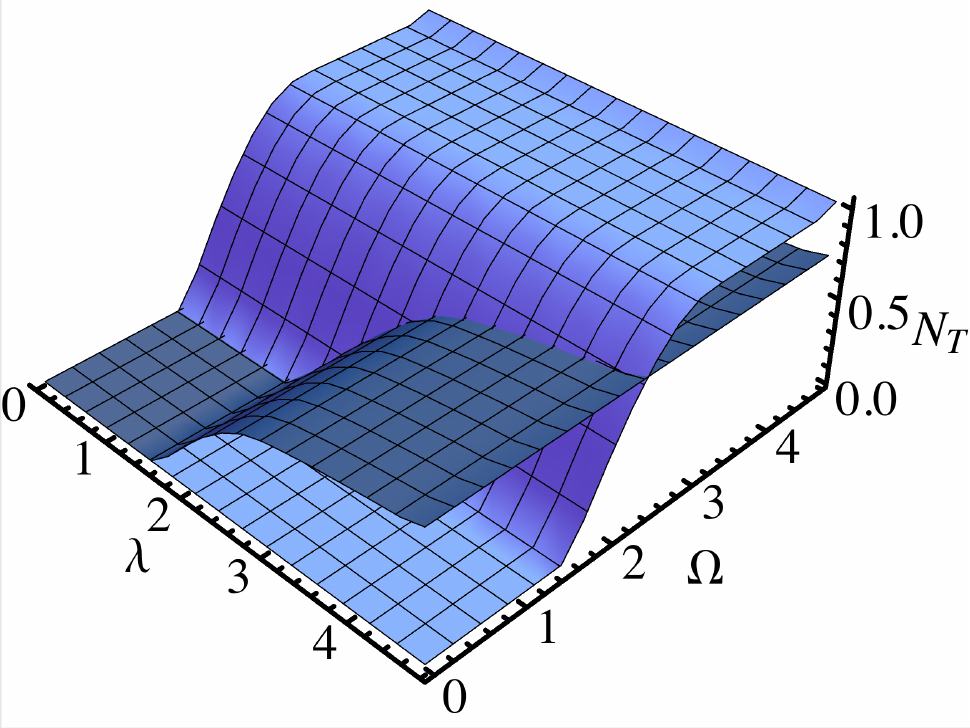}
\includegraphics[width=0.32\columnwidth]{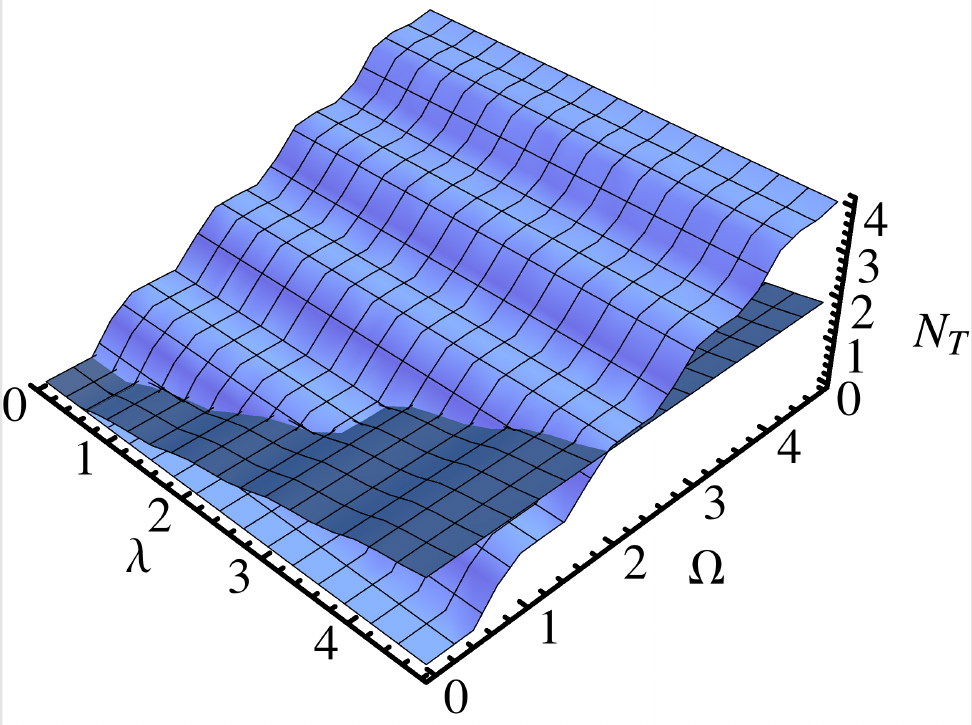}
\includegraphics[width=0.32\columnwidth]{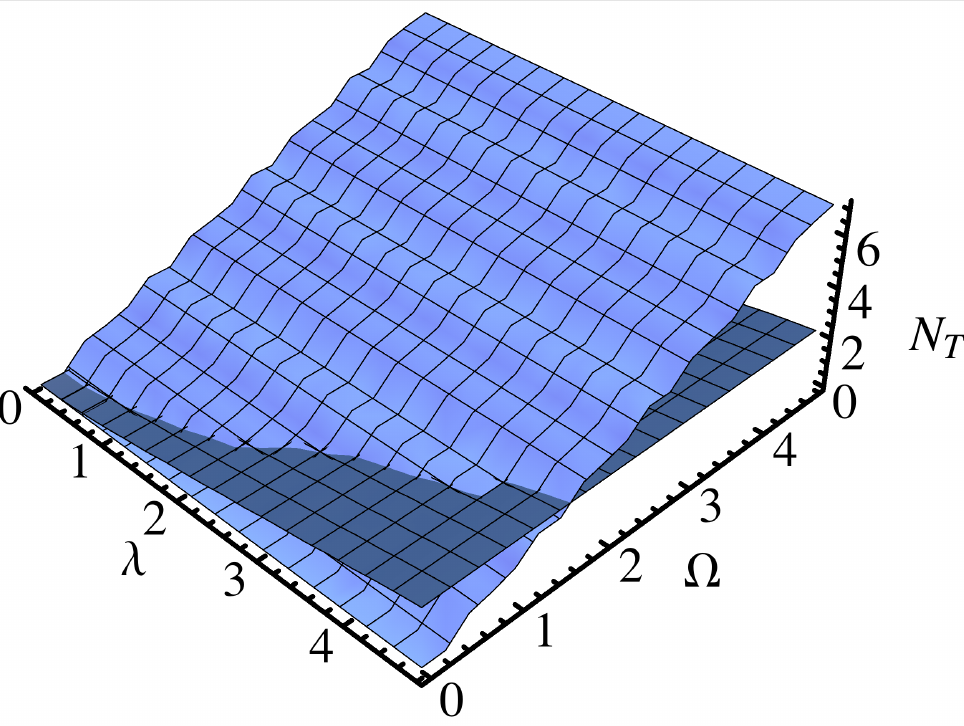}
\caption{Non-Markovianity $N_T$ versus $\lambda$ and $\Omega$ and 
for three different fixed values of ``length'' of the medium $T$. 
From left to right $T=1, 3, 5$. The two surfaces 
denote the integral in Eq. (\ref{nt}) in the $\Omega$-region, i.e. 
for $\theta=0$ (light gray) and in the $\lambda$-region, i.e. 
$\theta=\pi/2$ (dark gray). The non-Markovianity $N_T$ is the 
maximum between the two values. $N_T$ vanishes for small 
values of $\lambda$ and $\Omega$ and increases with both, 
as well as with the time $T$ (notice the different ranges on 
the $N_T$ axes).}
\label{Fig:1}
\end{figure}

Using the expression in Eq. (\ref{arhot}), the 
trace distance between a generic pair of evolved 
states may be written as follows
\begin{align}
D(\rho_{1t},\rho_{2t}) = \frac12 \sqrt{e^{-\gamma t}(W_{10}-W_{20})^2\cos^2  \lambda t + (M_{10}-M_{20})^2 \cos^2 \Omega t}\,.
\end{align}
The non-Markovianity measure $N_T$ corresponds to 
the maximal possible backflow of information, and is calculated 
by taking the maximum over all the initial pairs of states 
using Eq. (\ref{E20}). For qubit systems the maximum is obtained
for a pair of states that are  pure and orthogonal \cite{ref14,ref38}
and this means that we may write
\begin{align}
(W_{10}-W_{20})^2  = 4 \cos^2 \theta \qquad
(M_{10}-M_{20})^2  = 4 \sin^2 \theta \qquad \theta \in [0,\pi/2]\,. 
\notag
\end{align}
The maximisation over pairs of states is thus transformed into 
a maximisation over the single state parameter $\theta$, and the 
trace distance rate may be written as
\begin{align}
\sigma (t,\rho_{1t},\rho_{2t})  & \equiv \sigma (t,\gamma,\lambda,\Omega,\theta) \notag \\ & = -\frac{1}{2S} \left[e^{\frac12 \gamma t}  \cos^2 \theta\, \Omega \sin 2\Omega t + e^{-\frac12 \gamma t}\cos^2 \theta \left(
\gamma\sin^2 \lambda t + \lambda \sin 2 \lambda t \right)\right] \\
S & = \sqrt{e^{\gamma t}  \cos^2 \theta \cos^2 \Omega t + 
\sin^2 \theta \cos^2 \lambda t}\,.\notag
\end{align}
Before going to the explicit analysis of non-Markovianity, 
it is useful to notice a scaling property of $\sigma$, i.e. 
\begin{align}
\sigma (t,\gamma,\lambda,\Omega,\theta)
= \gamma\, \sigma (\gamma t,1,\lambda/\gamma,\Omega/\gamma,\theta)\,,
\end{align}
and, in turn, of $N_T$,
\begin{align}
N_T(\gamma,\lambda,\Omega) = N_{\gamma T}(1,\lambda,\Omega)\,.
\end{align}
As a consequence, one is led to consider the following 
dimensionless variables
\begin{align}
t \rightarrow \tau = \gamma t\,, \quad 
\Omega \rightarrow \Omega/\gamma \,, \quad
\lambda \rightarrow \lambda/\gamma \,, \quad
T \rightarrow \gamma T\,,
\end{align}
which amounts to measure all the considered quantities
in unit of $\gamma^{-1}$. With this choice one has 
\begin{align}
N_T \equiv N_T(\lambda,\Omega) = \max_\theta \int_{\sigma >0}^T \! 
d\tau\,  \sigma (\tau,1,\lambda,\Omega,\theta)\,.
\end{align}
Numerical analysis shows that for any given $T$, the maximum 
is obtained  either for $\theta=0$ or for $\theta  = \pi/2$, 
depending on the values of $\lambda$ and $\Omega$. Upon 
exploiting this result, we may write
\begin{align}
N_T(\lambda,\Omega) =  \int_{0}^T \! \!\!
d\tau \max\big[g_0(\tau,\Omega),g_{\pi/2} (\tau,\lambda)\big]\,,
\label{nt}
\end{align}
where
\begin{align}
g_0(\tau,\Omega) & = \frac12 
\Big[
\sigma (\tau,1,\lambda,\Omega,0) +
\big|\sigma (\tau,1,\lambda,\Omega,0)\big|
\Big] \notag \\ & = \frac{\Omega}{4} \frac{|\sin2\Omega\tau|-\sin2\Omega\tau}{|\cos\Omega\tau|},
\\ 
g_{\pi/2}(\tau,\lambda) & = \Big[
\sigma (\tau,1,\lambda,\Omega,\pi/2) +
\big|\sigma (\tau,1,\lambda,\Omega,\pi/2)\big|
\Big] \notag \\ &= \frac18 e^{-\tau/2}\, \frac{2\, \big|\cos^2 \lambda\tau + \lambda \sin 2\lambda \tau\big| - 1 - \cos 2\lambda \tau - 2\lambda \sin 2\lambda \tau}{|\cos\Omega\tau|}\,.
\end{align}
As a first consistency check, let us consider short time evolution, which should correspond to a Markovian behaviour. Indeed, we have
\begin{align}
\sigma (\tau,1,\lambda,\Omega,0) & \simeq - \Omega^2 \tau + O(\tau^2)\notag, \\
\sigma (\tau,1,\lambda,\Omega,\pi/2) & \simeq - \frac12 + \left(\frac14 - \lambda^2\right)\tau + O(\tau^2) \,,
\end{align} 
such that $N_T=0$. Using Eq. (\ref{nt}) we have then 
evaluated non-Markovianity as a function of  $\lambda$ and $\Omega$ 
for different values of the medium ``length'' $T$. Results are illustrated
in Fig. \ref{Fig:1}: as it is apparent from the plots, non-Markovianity 
is as an increasing function of  both $\lambda$ and $\Omega$. 
Moreover, for increasing $\Omega$, $N_T$ becomes almost independent
on $\lambda$. In Fig. \ref{Fig:1}, we show both the values obtained
for $\theta=0$ (light gray, referred to as the $\Omega$-region 
in the following) 
and $\theta=\pi/2$ (dark gray, $\lambda$-region) in order to
illustrate the regions in the $\lambda$-$\Omega$ plane where each
parameter is more relevant; remind that for $\theta=0$, $N_T$ does 
depend only on $\Omega$ and vice versa for $\theta=\pi/2$. The non-Markovianity is the maximum between the two values. Notice that
non-Markovianity increases with time, despite the dissipative 
nature of the interaction, and this fact suggests that the 
corresponding memory effects may be observed in practice. 
For sake of completeness we report the analytic expression of 
$N_T(\Omega)$ in the $\Omega$-region, the corresponding expression
in the $\lambda$-region is cumbersome and will not be reported here,
\begin{align}
N_T (\Omega) = \big[\big[\Omega T\big]\big] + \frac12 \left(\big|\cos \Omega T \big| 
- \big|\sin \Omega T\big| \cot \Omega T\right)\,,
\end{align}
where $[[x]]$ denotes the integer part of $x$. 
\par
\par
\begin{figure}[h!]
\includegraphics[width=0.32\columnwidth]{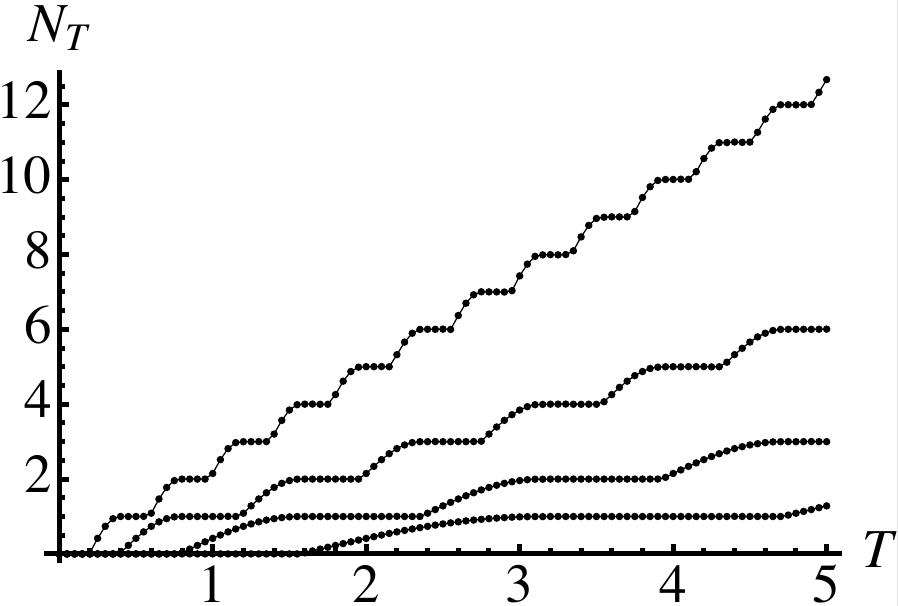}
\includegraphics[width=0.32\columnwidth]{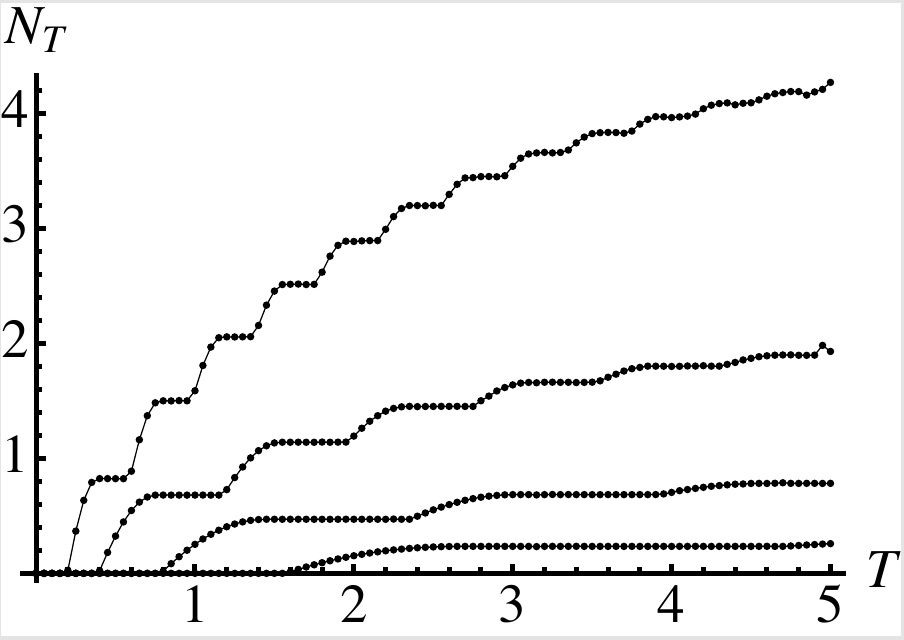}
\includegraphics[width=0.32\columnwidth]{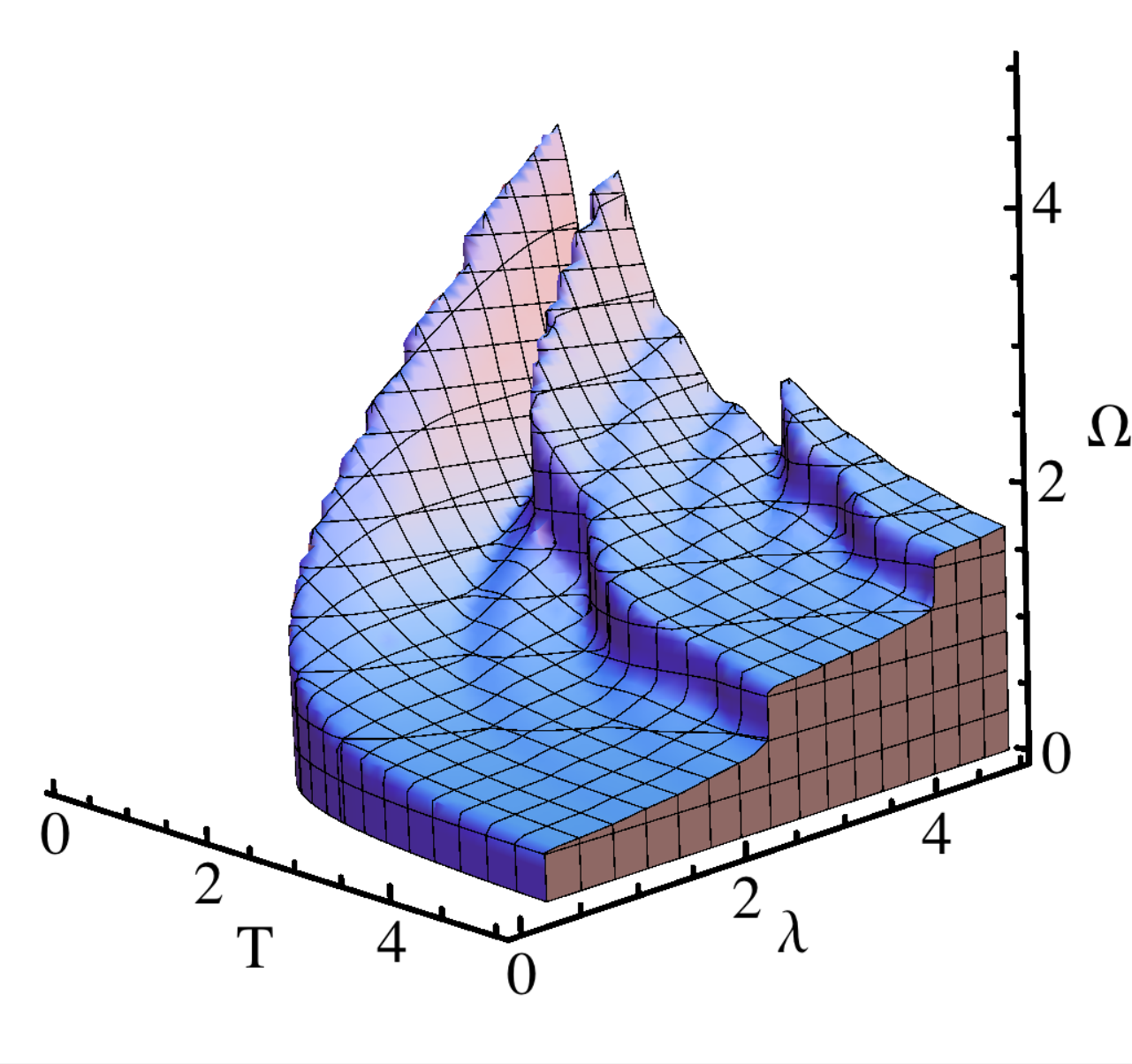}
\caption{Non-Markovianity $N_T$ as a function of $T$ for 
different values of $\lambda$ and $\Omega$. In the left 
panel we show $N_T$ vs $T$ for $\lambda=0.1$ 
and (from bottom to top) $\Omega=1, 2, 4, 8$, whereas in the central panel we show  $N_T$ vs $T$ for 
$\Omega=0.1$ and (from bottom to top) $\lambda= 1, 2, 4, 8$. 
Notice the different range for $N_T$ in the two panels.
The right panel shows the region in the 
$T-\lambda-\Omega$ parameter space where 
$\int_{0}^T \! d\tau\, g_{\pi/2} (\tau,\lambda) >
\int_{0}^T \! d\tau\, g_0 (\tau,\Omega)$ (in the range 
$T\in [0,5]$, $\lambda\in [0,5]$, $\Omega\in [0,5]$).}
\label{Fig:2}
\end{figure}

\noindent
In Fig. \ref{Fig:2} we show the behaviour of 
$N_T$ as a function of $T$ for different values of $\lambda$ and
$\Omega$. In the left panel we show $N_T$ vs $T$ for $\lambda=0.1$ 
and few values of $\Omega$, whereas in the central panel we show 
$N_T$ vs $T$ for $\Omega=0.1$ and few values of $\lambda$. As it is 
apparent from the plots, $N_T$ increases continuously, but not 
smoothly. Roughly speaking, $N_T$ is growing linearly with $T$ when 
$\Omega$ is the relevant parameter and sub-linearly viceversa. 
Notice the different range for $N_T$ in the two panels. In the right
panel we summarise the results, showing the region in the 
$T-\lambda-\Omega$ parameter space where 
$N_T =  \int_{0}^T \! 
d\tau\, g_{\pi/2} (\tau,\lambda) \equiv N_T (\lambda)$, i.e. where 
$\lambda$ is the relevant parameter determining the 
non-Markovianity. As mentioned before, this physically corresponds
to have a background field with a broad spectrum compared to the width of the transition.
Since random telegraph noise (RTN) is characterized by a 
Lorentzian spectrum, our results may be compared to those 
obtained for a two-level system interacting with a 
{\it dephasing} classical environment fluctuating according 
to RTN \cite{nmce}. It turns out that dipole interaction is 
leading to more pronounced non-Markovian effects compared 
to dephasing: backflow of information increases with time instead
of oscillating, and no threshold on the width of 
the spectrum (or the switching rate of the process) appears. 
This is a remarkable result, in view of the ubiquitous occurrence of 
the dipole interaction in nature. It may be also appropriate at
this point to remind that the non-Markovian character of a quantum
map resulting from the interaction with a classically fluctuating field is fully independent on the nature of the stochastic process used to
describe its time dependence \cite{smi11,apl}. 
In other words, the stochastic process
describing the field may be (classically) Markovian, 
whereas the quantum map originating from the interaction with the
corresponding classical field may be (quantum) non-Markovian at 
all times. This is exactly what happens for the telegraph noise,
which is described by a classically memoryless continuous-time 
stochastic process, but it leads to a non-Markovian quantum process
if the switching rate is small enough \cite{nmce}. 
\section{Conclusions}
In this paper, we have addressed the dynamics of a two-level 
system immersed in a classical fluctuating field, and interacting 
with its environment via dipole interaction. We have discussed 
the non-Markovianity of the corresponding quantum evolution, as 
measured by the  backflow of information (BLP measure) and, in 
particular, we have evaluated the backflow of information for 
a background field with a Lorentzian spectrum, also comparing 
our result with the analogue dephasing case.
\par
Our results uncovered the existence of two working regimes, which we
referred to as $\Omega$-regime or $\lambda$-regime, corresponding 
to sharp and broad spectrum of the field, where 
memory effects are governed either by the energy gap of the 
two-level system, or by the interaction energy, respectively.
Non-Markovianity vanishes for short interaction times, 
independently on $\lambda$ and $\Omega$ and then increases 
monotonically with time and with both $\lambda$ and $\Omega$, 
despite the dissipative nature of the 
interaction, thus suggesting that the corresponding memory 
effects may be observed in practical scenarios.
When compared to the corresponding dephasing dynamics, it 
turns out that dipole interaction is leading to more 
pronounced non-Markovian effects, 
with the backflow of information that increases with
the interaction time independently on width of the 
(Lorentzian) field spectrum. 
\section*{Acknowledgements}
DA was supported by Shahid Chamran University of Ahvaz [Grant No. 96/3/02/16670]. MGAP is member of GNFM-INdAM.
\section*{References}
\bibliographystyle{elsarticle-num} 

\begin{thebibliography}{00}
\bibitem {ref1}
H-P. Breuer and F. Petruccione, {\em The Theory of Open Quantum Systems}, (Oxford University Press, Oxford, 2002). 
\bibitem {ref2}
U. Weiss, {\em Quantum dissipative systms}, (World Scientific, Singapore, 2008). 
\bibitem {ref3}
H. J Carmichael and M. O Scully,  Phys. Today. {\bf 53}, 78 (2000). 
\bibitem {ref4}
G. Lindblad, Comm. Math. Phys. \textbf{48}, 119, (1976).
\bibitem {ref7}
S. Maniscalco and F. Petruccione, Phys. Rev. A \textbf{73},  012111, 
(2006).
\bibitem {ref8}
X.-T. Liang, Phys. Rev. E \textbf{82},  051918, (2010).
\bibitem {ref9}
C. W. Lai, P. Maletinsky, A. Badolato, and A. Imamoglu, 
Phys. Rev. Lett. \textbf{96}, 167403, (2006) .
\bibitem {ref5}
A. Chiuri, C. Greganti, L. Mazzola, M. Paternostro and P. 
Mataloni, Sci. Rep. \textbf{2}, 968, (2012).
\bibitem {ref6}
P. Lambropoulos, G. M. Nikolopoulos, T. R. Nielsen, and S. Bay, Rep. Prog. Phys. \textbf{63}, (2000) 455.
\bibitem{vas1} {R. Vasile, S. Olivares, M. G. A. Paris, S. Maniscalco}
Phys. Rev. A {\bf 83}, 042321 (2011).
\bibitem{add2} A. W. Chin, S. F. Huelga, M. B. Plenio
Phys. Rev. Lett. {\bf 109}, 233601 (2012).
\bibitem {ref10}
S. Nakajima, Progr. Theor. Phys. \textbf{20}, 968, (1958).
\bibitem {ref11}
R. Zwanzig, Physica \textbf{30}, 1109 (1964).
\bibitem {ref12} N. Hashitsume, F. Shibata, and 
M. Shingu, J. Stat. Phys. \textbf{17}, 155 (1977); 
F. Shibata, Y. Takahashi, and N. Hashitsume, J. Stat. Phys. 
\textbf{17}, 171 (1977).
\bibitem {ref13}
S. Chaturvedi and F. Shibata, Z. Phys. B \textbf{35}, (1979).
\bibitem {ref14}
E.-M. Laine, J. Piilo and H.-P. Breuer, Phys. Rev. A 
\textbf{81}, 062115, (2010).
\bibitem {ref15}
A. Rivas, S. F. Huelga, and M. B. Plenio, Phys. Rev. Lett. \textbf{105},  050403, (2010).
\bibitem {ref16}
X.-M. Lu, X. Wang and C. P. Sun, Phys. Rev. A \textbf{82}, 042103, (2010).
\bibitem {ref17}
S. Luo, S. Fu and H. Song, Phys. Rev. A \textbf{86},  044101, (2012).
\bibitem {ref18}
S.-L. Chen \textit{et al.},  Phys. Rev. Lett. {\bf 116}, 020503,  (2016).
\bibitem {ref19}
C. Addis, B. Bylicka, D. Chru\`sci\`nski, and S. 
Maniscalco, Phys. Rev. A \textbf{90},  052103, (2014).
\bibitem {ref20}
T. J. G. Apollaro, C. Di Franco, F. Plastina and M. Paternostro, Phys. Rev. A \textbf{83},  032103, (2011).
\bibitem {ref21}
Z. Xu, W. Yang, and M. Feng, Phys. Rev. A \textbf{81},  044105, (2010).
\bibitem {ref22}
Y.-J. Zhang, W. Han, Y.-J. Xia, J.-P. Cao, and H. 
Fan, Sci. Rep. \textbf{4},  4890,  (2014).
\bibitem {ref23}
B. Bellomo, G. Compagno, R. L. Franco, A. Ridolfo, S. Savasta, 
Phys. Scripta T\textbf{147},  014004, (2012).
\bibitem {ref24}
C. Pineda, T. Gorin, D. Davalos, D. A. Wisniacki, and I. 
Garcia-Mata, Phys. Rev. A \textbf{93},  022117, (2016).
\bibitem {ref25}
Z. X. Man, N. B. An, and Y. J. Xia, Phys. Rev. A \textbf{90},  062104,
(2014).
\bibitem {ref26}
A. Smirne, L. Mazzola, M. Paternostro and B. Vacchini, Phys. Rev. A \textbf{82},  062114, (2010).
\bibitem {ref27}
S. Alipour, A. Mani and A. T. Rezakhani, Phys. 
Rev. A \textbf{85},  052108, (2012).
\bibitem {ref28}
H.-S. Zeng, N. Tang, Y.-P. Zheng, and G.- Y. Wang, 
Phys. Rev A \textbf{84},   032118,(2011).
\bibitem {ref33}
M. Mannone, R. Lo Franco and G. Compagno, Phys. Scr. 
T\textbf{153},  014047, (2013).
\bibitem {ref30}
Y. J. Zhang, Y. J. Xia, and H. Fan,  EPL \textbf{116},  30001, 
(2016).
\bibitem {ref34}
J.-G. Li, J. Zou, and B. Shao, Phys. Rev. A \textbf{81},  062124,(2010).
\bibitem {ref35}
F. W. Cummings, Am. J. Phys. \textbf{30}, 898, (1962).
\bibitem {ref36}
F. W. Cummings, Nuovo Cim. B \textbf{70}, 102, (1982).
\bibitem {oli12}
F. Benatti, R. Floreanini, S. Olivares,
Phys. Lett. A {\bf 376}, 2951 (2012). 
\bibitem{nmce} C. Benedetti, M. G. A. Paris, S. Maniscalco,
Phys. Rev. A {\bf 89}, 012114 (2014).
\bibitem{qpFGN} M. G. A. Paris, Physica A {\bf 413}, 256 (2014).
\bibitem {ref38}
S. Wissmann, A. Karlsson, E.-M. Laine, J. Piilo, and H.-P. Breuer, Phys. Rev. A \textbf{86},   062108, (2012).
\bibitem{smi11}
B. Vacchini, A. Smirne, E.-M. Laine, J. Piilo and H.-P. Breuer
New J. Phys. {\bf 13}, 093004 (2011).
\bibitem{apl} S. Cialdi \textit{et al.},
Appl. Phys. Lett.  \textbf{110}, 081107 (2017).
\end{thebibliography}

\end{document}